\documentstyle[prb,aps,epsf]{revtex}
\begin{document}
\twocolumn[
\hsize\textwidth\columnwidth\hsize\csname @twocolumnfalse\endcsname
\title{Steric constraints in model proteins}
\author{Cristian Micheletti$^1$, Jayanth R. Banavar$^2$, Amos
Maritan$^3$ and
Flavio Seno$^1$}
\vskip 0.3cm
\address{(1) Istituto Nazionale per la Fisica della
Materia \\ Dipartimento di Fisica, Universit\`a di Padova, 
Via Marzolo 8,35131  Padova, Italy }

\address{(2) Department of Physics and Center for Materials Physics,
104 Davey Laboratory, \\ The Pennsylvania State University, University
Park, Pennsylvania 16802}

\address{(3) Istituto Nazionale per la Fisica della
    Materia  \\ International School for Advanced Studies (S.I.S.S.A.),
Via Beirut 2-4, 34014 Trieste, Italy }

\date{November 18, 1997}
\maketitle
\begin{abstract}
A simple lattice model for proteins that allows for distinct sizes of
the amino acids is presented.  The model is found to lead to a
significant number of conformations that are the unique ground state
of one or more sequences or encodable. Furthermore, several of the
encodable structures are highly designable and are the non-degenerate
ground state of several sequences.  Even though the native state
conformations are typically compact, not all compact conformations are
encodable.  The incorporation of the hydrophobic and polar nature of
amino acids further enhances the attractive features of the model.

\noindent {Pacs numbers: 87.10.+e, 87.15.By, 64.60.CN, 61.41.+e.}
\end{abstract}
\vskip 0.3cm
]

Protein folding remains a major unsolved problem in molecular biology.
The successful design of proteins and enzymes with desired
functionality and native state structure requires a knowledge of the
rules underlying protein architecture \cite{1}. Simple exact models
have proved to be invaluable in deducing the general principles of
protein structure and stability \cite{2,3,4}. Such models provide a
coarse-grained description of proteins and have provided crucial
insights on the role of inter-amino acid interactions (and in
particular the hydrophobicity \cite{3} of some residues) in
determining the secondary and tertiary structures of proteins. An even
more fundamental structural principle pertains to steric constraints
related to the diversity of residue sizes, the prohibition of overlaps
of atoms and a close packing of the residues leading to small cavity
volumes \cite{5}. The principal theme of this paper is the
introduction of a simple coarse-grained lattice model of a protein
that consists of a sequence of amino acids configured in a
self-avoiding conformation and which takes into account these steric
constraints.

In its simplest form, the new model (that we will denote as the $LS$
model) has two types of amino acids denoted by $L$ (large) and $S$
(small).  (Further refinements taking into account a range of residue
sizes is straightforward in principle but not essential to capture the
role of steric constraints in determining protein architecture.)  In
terms of their interactions, the model is merely that of a
homopolymer, a polymer made up of identical units, with an attractive
energy proportional to the number of contacts between amino acids.
Two amino acids are said to be in contact when they sit on adjoining
sites but are yet not next to each other in sequence.  Thus the lowest
energy conformation of the sequence is one that has the largest
possible number of contacts allowed by the steric interactions.
Furthermore, if a particular sequence is able to be configured in a
maximally compact conformation (which has the largest possible number
of contacts and therefore is one of lowest energy), then any other
degenerate ground state must also necessarily be maximally compact.

The consideration of steric constraints arises from postulating that
in order to accomodate the large size of an $L$ amino acid, at least
one of the sites next to it must be kept vacant -- no such constraint
is imposed on the $S$ amino acid.  The steric constraint on the
$L$-type amino acids immediately rules out the possibilty of any
maximally compact conformation in which $L$ amino acids sit in the
interior.  Thus any allowed maximally compact conformation (which as
stated above has an energy that cannot be improved upon) will have
$S$-type amino acids in the interior and either $S$ or $L$-type amino
acids at the surface.  We have carried out extensive tests of the $LS$
model on a two dimensional square lattice and have found that it has
many of the desirable attributes for modelling proteins:

\begin{enumerate}
\item {the lowest energy states are typically compact;}
\item {only a tiny fraction of sequences admit a unique ground
    state (to ensure specificity);}
\item {there exist a significant number of encodable
    structures, i.e. structures which are the unique ground state of
    one or more sequences;}
\item {some of the  encodable structures have a high
    degree of encodability and are thus highly designable, i.e.
    they are the non-degenerate  ground state of
    several sequences.}
\end{enumerate}

Perhaps the simplest existing lattice model that satisfy these
criteria is the $HP$ model of Lau and Dill \cite{3} which also
considers two kinds of amino acids denoted by $H$ (hydrophobic) and
$P$ (polar).  In the $HP$ model, which has been studied widely, the
protein collapse is driven by hydrophobic interactions between the $H$
amino acids and the solvent.  After integrating the solvent degrees of
freedom, a more attractive effective $H-H$ interaction compared to the
effective $P-P$ and $H-P$ interactions results.  Thus, in its native
state conformation, which is typically compact, the interior residues
are usually hydrophobic and are thus shielded from the solvent.  As a
benchmark for our results, we will often refer to comparable results
in the $HP$ model with an attractive $H-H$ interaction and zero $H-P$
and $P-P$ interactions.

The $LS$ model may be formally described by the Hamiltonian:

\begin{equation}
{\cal H} = - \sum_i [z_i(\Gamma)] \cdot A[z(\sigma_i) -z_i(\Gamma)]
\label{eqn:ham}
\end{equation}

\noindent where $\sigma_i \in \{L,\ S \}$, $z(\sigma_i)$ is the number
of bonds belonging to residue $i$ not used for chain connectivity and
on a square lattice:

\begin{equation}
z(\sigma_i) = \left\{
  \begin{array}{l}
   1 \ \ \ \mbox{for $L$ residues inside the chain,}\\ 
   2 \ \ \ \mbox{for $S$ residues inside the chain,}\\ 
   2 \ \ \ \mbox{for $L$ residues at chain ends,}\\ 
   3 \ \ \ \mbox{for $S$ residues at chain ends.}
  \end{array}
\right.
\end{equation}

$z_i(\Gamma)$ is the number of contacts of the $i$th residue in a
conformation $\Gamma$
and $A(x)$ is defined by

\begin{equation}
A(x) = \left\{
  \begin{array}{l}
        1 \ \ \mbox{if $x \ge 0$,}\\
        -\infty \ \ \mbox{otherwise.}   
  \end{array}
\right.
\end{equation}

\noindent The function $A(x)$ is used to enforce residue
incompressibility. In fact, when mounting a sequence on a structure,
it may happen that a $L$ residue is surrounded by four occupied
sites. In this case $A[z(\sigma_i) -z_i(\Gamma)]$ diverges, thus
assigning a $+\infty$ energy penalty to this forbidden situation.  The
model may be generalized in a straightforward way to higher dimensions
and variations of sizes and the nature of steric constraints.  The
latter may be softened by allowing for a variety of possibilities but
with an associated cost, for example by modifying the definition of
$A(x)$.

We now describe the results of our exact enumeration studies on
two-dimensional square lattices. Chan and Dill \cite{3} have shown that two
dimensional models more faithfully capture the correct physically
important surface-interior ratios of proteins than the corresponding
three dimensional counterparts.  They point out that in order to
reproduce the correct ratio for a molecule of the size of myoglobin
requires only 16-20 monomers in two dimensions as opposed to 154 in
three dimensions. The latter case is clearly beyond the scope
of exact enumeration. A more feasible size in three dimensions would
consist of 27 monomers, but unfortunately has just one interior
residue in a maximally compact conformation.

The core of the computational approach is made up of two backtracking
procedures through which one generates the complete set of sequences
and inequivalent structures of a given length -- one exploits lattice
symmetries to get rid of redundant structures.  Following standard
practice, we will assume that head to tail inversions is not an
allowed symmetry operation and that we are dealing with oriented
walks.  We have carried out a complete enumeration of all sequences of
length 16 and all self-avoiding conformations for both the $LS$ and
$HP$ models.  Altogether there are $2^{16}=65536$ different sequences
and 802075 distinct oriented walks (69 of which are maximally compact
and fill a 4x4 square). We laid out each sequence on each of the
802075 structures and determined its ground states. We kept track of
both the sequences which happened to have a unique ground state and
also its native conformation.  For the $LS$ model, the total number of
sequences with a unique ground state was 7555, while the number of
distinct encodable structures was 117. For each of these structures we
calculated its encodability score, i.e., the number of sequences that
admit it as the unique ground state.  A summary of our results is
presented in Table I.

It is striking that the most encodable structures for the $LS$ model
are maximally compact (Figure 1).  Indeed there are a grand total of
7202 sequences (out of 7555) that admit a maximally compact structure
as their unique native states. It is also important to note that not
all maximally compact structures are encodable (thus preserving
specificity). In fact, only 33 out of the 69 maximally compact
structures are encodable.  The encodability scores of these compact
structures range from 60 up to 519.  The $HP$ model, on the other
hand, admits 456 encodable structures and the highest encodability
score is 26. Only 20 of these structures turn out to be maximally
compact.

We have also carried out exact enumeration studies on chains of length
16 considering only the 69 maximally compact structures as target
ground states. For the $HP$ model, due to the absence of significant
competing structures which are non-compact, all 69 conformations are
recognized as encodable, with an associated loss of specificity. On
the contrary the encodable maximally compact structures of the $LS$
model are not affected. This property can be rigorously established by
geometrical arguments, as well as the fact that, if a sequence admits
a unique compact structure as a ground state, then it cannot be mounted on
any other compact conformation without violating steric constraints.
This implies an enhanced thermodynamic stability of encodable compact
structures with respect to the $HP$ model. In fact, the average energy
gap of encodable structures measured on the compact ensemble is
infinite for the $LS$ model whereas it is finite for the $HP$ case
\cite{4}.

We now turn to the case of sequences of length 25. The number of
distinct structures of length 25 is too big to allow an exhaustive
search for each of the $2^{25}$ sequences.  However, because it
appears that the most significant structures are the maximally compact
ones, we simplified the task by considering only such structures as
candidate ground states for each of the $2^{25}$ sequences.  For the
$LS$ model, the number of sequences of length 25 that admit a unique
ground state on one of the 1081 compact structures is 1340155. The
number of encodable compact structures is 589 (Table I and Figure 2).
It is important to note that these numbers are rigorous lower bounds
to the number that would be obtained, were one to consider the
non-compact conformations as well.  This follows from the observation
that a maximally compact conformation has a lower energy than any
non-compact conformation for the $LS$ model.

The case of chains of 36 beads is interesting because they are
sufficiently long to reveal the presence or absence of geometrical
regularity in highly encodable structures.  Again, to reduce the
numerical task to manageable proportions, we will consider only the
maximally compact structures that fit in a 6x6 square. There are
altogether 57337 inequivalent compact structures. It is not
numerically feasible to mount each of the $2^{36}$ $LS$ sequences on
the whole set of compact structures to determine whether they have a
unique ground state.  We therefore chose to explore a tiny portion of
the sequence space by means of random sampling. By using a good random
number generator \cite{6}, we generated 12996000 random $LS$
sequences.  We found that 16611 of the compact structures were
encodable with the encodability scores ranging between 1 and 64.
Figure 3 shows pictures of the inequivalent structures with the top
encodability scores and show motifs of emergent secondary structure.
Interestingly, the most encodable structure is the same as the most
designable structure found by Li {\em et al.} \cite{4} within the HP
framework (for parameters $E_{HH}=-2.3$, $E_{HP}=-1.0$, $E_{PP}=0.0$).

{} From the results presented so far, it appears that the $LS$ model
captures encodability and specificity better than the $HP$ model
suggesting that steric constraints could be as important as
hydrophobic/polar interactions in determining protein architecture.
An interesting avenue for further exploration would be a study of the
kinetics of folding of the $LS$ model and comparisons with the $HP$
model.  Another intriguing possibility is to combine the two classes
($LS$ and $HP$) with the introduction of four species of residues:
Large-Hydrophobic ($LH$), Large-Polar ($LP$), Small-Hydrophobic ($SH$)
and Small-Polar ($SP$). Would one end up with a larger number of
encodable structures or higher encodability scores? Would this cure
the well-known defect of the pure $HP$ model, namely that $HP$
sequences have, on average, a huge ground-state degeneracy?

An exact statement that one may make is that if one considers $LS$
sequences that have a unique maximally compact ground state, then on
introducing an additional $H$ or $P$ character, the same ground state
will be retained unless the new ground state of the $LSHP$ model is no
longer maximally compact.  In order to address this issue
quantitatively, we have carried out an exact enumeration of the
$4^{12}$sequences of length 12 on a square lattice.  Our results
(Table I) show that the $HP$ diversity stabilizes new encodable
structures (though {\bf not} just compact ones!). Also note that there are
many ways of choosing how to ``fine-grain'' the original $LS$ sequence
into a $LSHP$ one (or equivalently many ways of ``coloring'' the $LS$
residues in order to assign them a hydrophobic/polar index).  These
observations are consistent with the idea that the introduction of a
moderate diversity of the complementary kind may, indeed, improve the
"good features" of the pure $LS$ or $HP$ models by enhancing the
encodability of designable compact structures. It is also expected, on
general grounds, that the increased amino acid diversity will cause
sequences to have, on average, a smaller ground state degeneracy than
in the pure $HP$ lattice model.

Our studies show that the most encodable structures are indeed compact
and that the smaller residues are more easily accomodated within the
core of the protein.  In order to minimize the energy of the native
state of the protein, hydrophobic amino acids tend to be buried within
the core (in order to avoid the solvent).  Another way of enhancing
the thermodynamic stability of the native state is by increasing the
energies of the sequence in competing conformations \cite{7} and may
result in polar amino acids being found in the core of the native
state.  The considerations presented here suggest that the smaller
polar amino acids (Thr, Ser) ought to have a larger propensity for
being buried than the larger ones (Lys, Glu and Arg) and this is
indeed observed in studies of natural proteins \cite{8}.

{ \bf Acknowledgments:} We are indebted to Gautham Nadig for helpful
discussions.  This work was supported in part by INFN sez. di Trieste,
NASA, NATO and the Center for Academic Computing at Penn State.

\begin{center} \begin{table} \begin{tabular}
{|| c| c | c | c| c ||} \hline \hline 
Model & N & ES & CES & MDS \\ 
\hline \hline 
LS & 12 & 15 & 15 & 24 \\ 
HP & 12 & 25 & 5 & 14 \\ 
LSHP & 12 & 232 & 31 & 4 $\cdot 10^5$\\   
LS & 16 & 117 & 33 & 519 \\ 
HP & 16 & 456 & 20 & 26 \\ 
LS & 25 & --- & 589 & 12777 \\
\hline \hline 
\end{tabular} 
\caption{The chain length $N$, the number
  of Encodable Structures (ES), the number of Compact Encodable
  Structures (CES) and the Maximum Designability Score (MDS)
  on a square lattice.  For $N=25$, only maximally compact
  conformations were considered.}
\end{table}
\end{center}

\newpage

\begin{figure}[hb]
\centerline{
 \epsfxsize=4.5in
 \epsffile{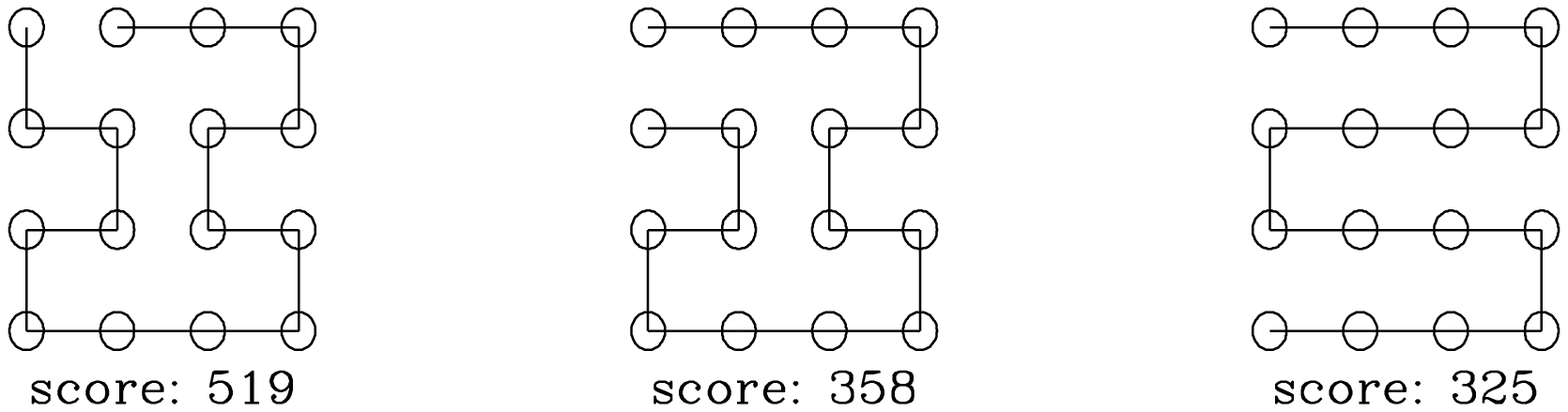}
 }
\caption[1a]{\label{fig1} The most encodable compact structures not
related by any symmetry operation (with their designability score) for
the $LS$ model in $d=2$ with $N=16$ monomers. The data are obtained
with an exact enumeration of all the sequences and all the
conformations.}
\end{figure}

\begin{figure}[hb]
\centerline{
 \epsfxsize=3.5in
 \epsffile{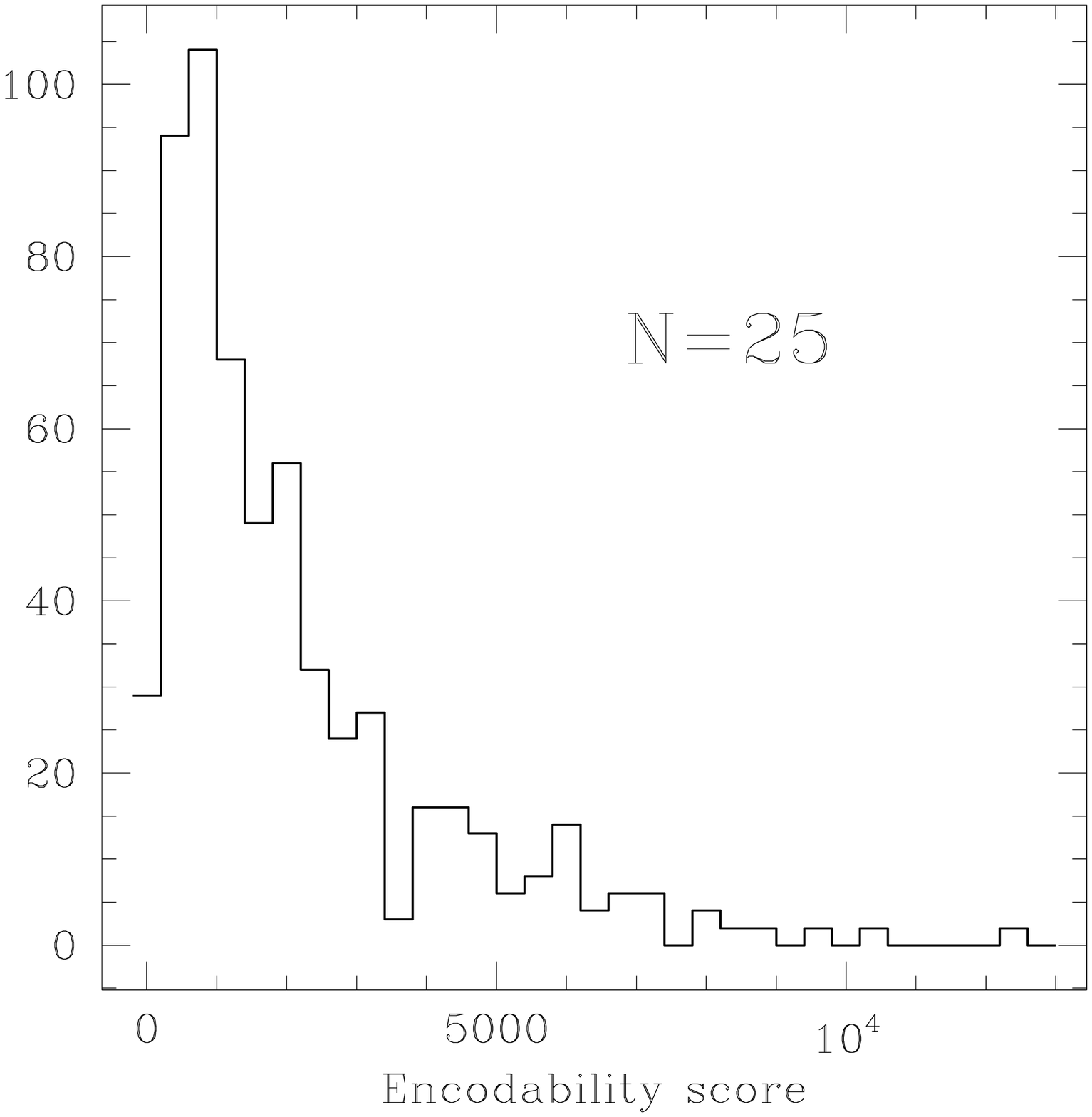}
 }
\caption[2]{\label{fig2} Histogram of number of compact structures  with a
given encodability score for the $LS$ model with $N=25$. }
\end{figure}

\begin{figure}[hb]
\centerline{
 \epsfxsize=3.5in
 \epsffile{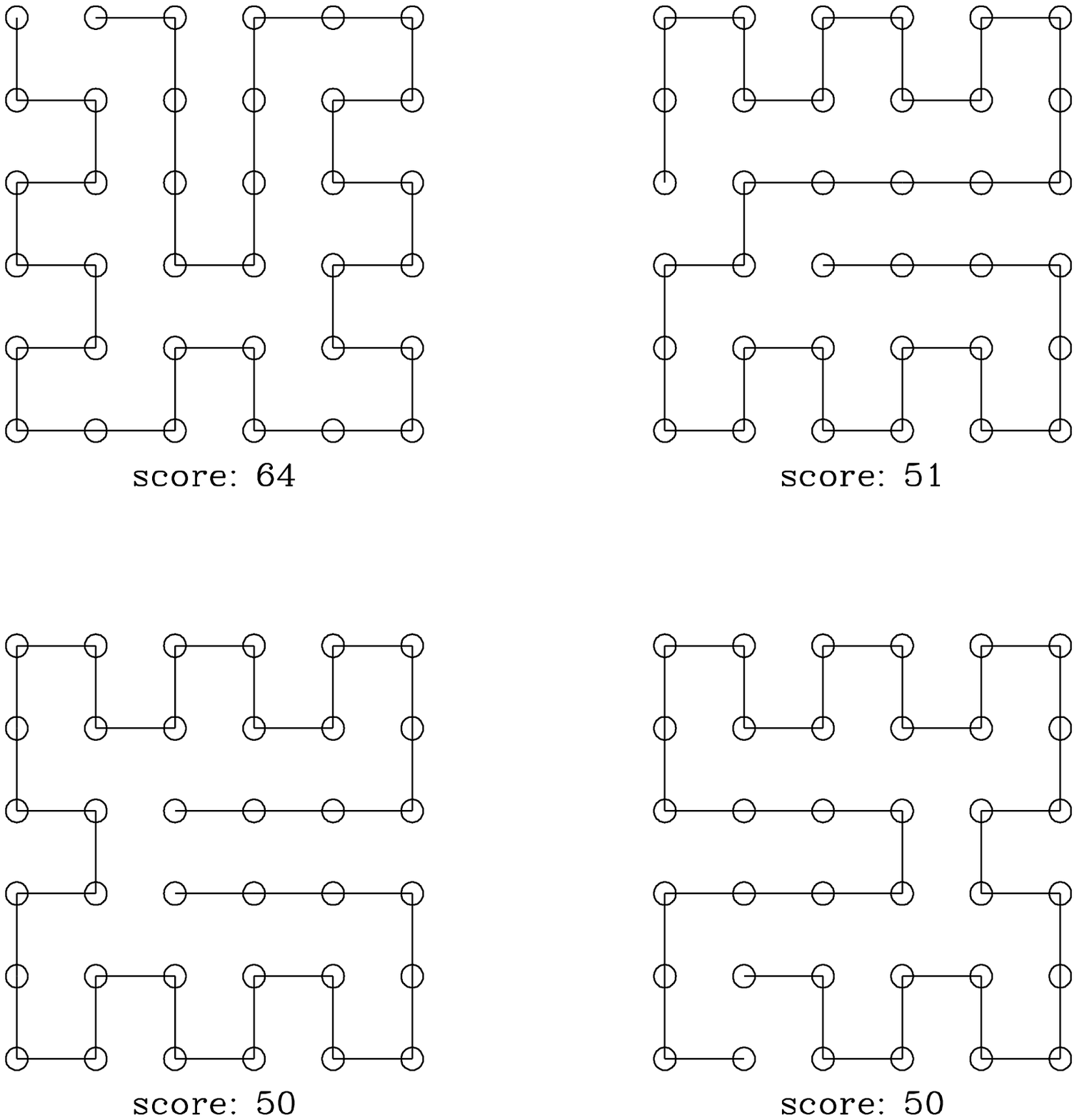}
 }
\caption[2]{\label{fig3} The most encodable compact structures (with
 their score) for the $LS$ model with $N=36$. The results are obtained
 with a random sampling of the space of the sequences and a
 complete enumeration of all the compact structures.}
\end{figure}


\begin{references}
\bibitem{1} C. Branden, \& J. Tooze, (1991) in {\em Introduction to
protein structure}, Garland Publishing, New York; T.E. Creighton,
(1983) in {\em Proteins: structures and molecular properties},
W. H. Freeman ed., New York
\bibitem{2} See, e.g., C. J. Camacho and D. Thirumalai, {\em
Proc. Nat. Acad. Sci. U.S.A.}  {\bf 90}, 6369 (1993); N. D. Socci and
J. N. Onuchic, {\em J. Chem. Phys.} {\bf 101}, 1519 (1994); A. Sali,
E. Shakhnovich and M. Karplus, {\em Nature}, {\bf 369}, 6477 (1994);
M.R. Betancourt and J.N. Onuchic, {\em J. Chem. Phys.}, {\bf 103}, 773
(1995); J. N. Onuchic, P. G. Wolynes, Z. Luthey-Schulten and
N. D. Socci, {\em Proc. Nat. Acad. Sci.} {\bf 92}, 3626 (1995);
J.M. Deutsch and T. Kurosky, {\em Phys. Rev. Lett.} {\bf 76}, 323
(1996); F. Seno, M. Vendruscolo, A. Maritan and J. R. Banavar, {\em
Phys. Rev. Lett.}  {\bf 77}, 1901 (1996); D. K. Klimov and
D. Thirumalai, {\em Phys. Rev. Lett.}, {\bf 76}, 4070 (1996); L. Mirny
and E. Domany, {\em Proteins: Structure, Function and Genetics}, {\bf
26}, 391 (1996); M.P. Morrissey and E.I. Shakhnovich, {\em Folding and
Design}, {\bf 1}, 229 (1996); L.A. Mirny and E.I. Shakhnovich, {\em
J. Mol. Biol.}, {\bf 264}, 1164 (1996); M. Cieplak, S. Vishveshwara
and J. R. Banavar, {\em Phys. Rev. Lett.} {\bf 77}, 3681 (1996);
M.H. Hao and H.A. Scheraga,{\em Physica }, {\bf A244}, 124 (1997);
H. Li, C. Tang and N.S. Wingreen, {\em Phys. Rev. Lett.}, {\bf 79},
765 (1997).
\bibitem{3} K. F. Lau and K. A. Dill, {\em Macromolecules} {\bf 22},
3986-3997 (1989); H.S.  Chan and K. A. Dill, {\em Physics Today}, {\bf
46}, 24 (1993); K. A. Dill, S.  Bromberg, S. Yue, K.  Fiebig, K. M.
Yee, P. D. Thomas and H. S. Chan, {\em Protein Science} {\bf 4}, 561
(1995); P.D. Thomas and K.A. Dill, {\em Proc. Natl. Acad.  Sci. USA},
{\bf 93}, 11628 (1996).
\bibitem{4} H. Li, R. Helling, C. Tang and N. Wingreen {\em Science},
{\bf 273}, 666 (1996).
\bibitem{5} G. N. Ramachandran and V. Sasisekharan, {\em
Advan. Prot. Chem.} {\bf 23}, 283 (1968); J. W. Ponder and
F. M. Richards, {\em J. Mol. Biol.}  {\bf 193}, 775 (1987).
\bibitem{6} S. Kirkpatrick and E. Stoll, {\em J. Comp. Phys.} {\bf
40}, 517 (1981).
\bibitem{7} See, e.g., J. R. Banavar, M. Cieplak, A. Maritan,
G. Nadig, F. Seno and S. Vishveshwara, {\em Proteins: Structure,
Function and Genetics} (in press).
\bibitem{8} C. Chothia, {\em J. Mol. Biol.} {\bf 105}, 1 (1976); J.
Janin, {\em Nature}, {\bf 277}, 491 (1979); G. D. Rose,
A. R. Gesolowitz, G. J. Lesser, R. A. Lee and M. H. Zehfus, {\em
Science} {\bf 229}, 834 (1985); S. J. Hubbard, K. H. Gross and
P. Argos, {\em Prot. Engg.} {\bf 7}, 613 (1994).
\end{references}
\end{document}